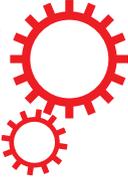

www.nature.com/scientificreports

# SCIENTIFIC REPORTS

OPEN

# Fast Maximum Likelihood Estimation via Equilibrium Expectation for Large Network Data

Maksym Byshkin[1], Alex Stivala[2,1], Antonietta Mira[1,3], Garry Robins[4,2] & Alessandro Lomi[1,4]

A major line of contemporary research on complex networks is based on the development of statistical models that specify the local motifs associated with macro-structural properties observed in actual networks. This statistical approach becomes increasingly problematic as network size increases. In the context of current research on efficient estimation of models for large network data sets, we propose a fast algorithm for maximum likelihood estimation (MLE) that affords a significant increase in the size of networks amenable to direct empirical analysis. The algorithm we propose in this paper relies on properties of Markov chains at equilibrium, and for this reason it is called equilibrium expectation (EE). We demonstrate the performance of the EE algorithm in the context of exponential random graph models (ERGMs) a family of statistical models commonly used in empirical research based on network data observed at a single period in time. Thus far, the lack of efficient computational strategies has limited the empirical scope of ERGMs to relatively small networks with a few thousand nodes. The approach we propose allows a dramatic increase in the size of networks that may be analyzed using ERGMs. This is illustrated in an analysis of several biological networks and one social network with 104,103 nodes.

Developing efficient approaches to data analysis and statistical inference is becoming increasingly important due to the widespread availability of large data sets in many fields of science. This is particularly the case for relational data typically taking the form of square arrays recording the presence of one or more relations among units of interest represented as network nodes[1,2]. Network representation of data is a fundamental tool for understanding and modeling a wide range of complex systems. Among the available models for relational data, exponential random graph models (ERGMs) are generally considered as "The most promising class of statistical models for expressing structural properties of social networks observed at one moment in time"[3]. ERGMs have found broad application in the analysis of social networks[4], as well as biological networks[5].

Formally, ERGMs may be written in the following form:

$$\pi(x, \boldsymbol{\theta}) = \frac{1}{k(\boldsymbol{\theta})} \exp\left(\sum_A \theta_A z_A(x)\right), \qquad (1)$$

which expresses the probability of observing a network with a fixed number nodes in a given state $x$. Here $z_A(x)$ are network statistics from the space of states $x$, which are counts of theoretically meaningful or empirically relevant network subgraphs (often called "configurations" in ERGM terminology). The summation is over all configurations $A$, $\theta_A$ denotes the model parameter associated to $z_A(x)$, $\boldsymbol{\theta}$ is the vector of all these parameters and $k(\boldsymbol{\theta})$ the normalizing constant ensuring that the probability distribution sums to one. The reader may also recognize equation (1) as an exponential family in canonical form (see Ch. 8 of Barndorff-Nielsen[6]), a Gibbs distribution, or a Markov random field[7].

[1]Institute of Computational Science, Università della Svizzera italiana, Lugano, 6900, Switzerland. [2]Centre for Transformative Innovation, Swinburne University of Technology, Hawthorn Victoria, 3122, Australia. [3]Dipartimento di Scienza e Alta Tecnologia, Università dell'Insubria, Como, 22100, Italy. [4]School of Psychological Sciences, University of Melbourne, Parkville Victoria, 3010, Australia. Correspondence and requests for materials should be addressed to A.L. (email: alessandro.lomi@usi.ch)

Different structural features are present in different networks. Yet, empirical network data are also characterized by recurrent structural regularities whose identification is crucial for understanding the behavior of complex network systems. For instance, much recent research has concerned "motifs," small subgraphs occurring more frequently than might be expected by chance (and hence often similar to ERGMs configurations)[8,9]. Motifs have been considered as the building blocks of complex networks. To determine if a given motif is over-represented, the frequency of the motif in an observed network is typically compared to the average frequency in an appropriate random network (null model)[10–12]. ERGMs permit inference on the under or over-representation of specific configurations *conditional* on the presence of other configurations, or structural features. In ERGMs, network statistics may be defined for particular subgraphs that may be of general or contingent interest in the particular network under study like, for example, reciprocity or triadic closure[3,4,13–15]. Network nodes may have attributes[16], the expression for $z_A(x)$ may be rather complicated, and the number of parameters may be large. Common examples of network statistics adopted in empirical research may be found elsewhere[3,4,17].

The model parameters fit the observed network $x_{obs}$ if the following method of moments (MoM) condition[18] is satisfied for all $A$:

$$E_{\pi(\theta)}(z_A(x)) = z_A(x_{obs}), \qquad (2)$$

where $E_{\pi(\theta)}(z_A(x)) = \sum_x z_A(x)\pi(x, \theta)$ denotes the expected network statistics with respect to the probability distribution (1) with parameters $\theta$ to be estimated; $z_A(x_{obs})$ denotes network statistics in the observed network. If the estimated $\theta_A$ is significantly larger than zero, then the corresponding network statistic $z_A$ occurs more frequently than might be expected by chance given all the other parameters of the model, and the corresponding configuration is over-represented. A well-known property of the exponential family (e.g., Ch. 8 of Barndorff-Nielsen[6]) is that $E_{\pi(\theta)}(z_A(x))$ is a monotonically increasing function of $\theta_A$. Thus, the estimated $\theta_A$ measures the corresponding structural features in observed networks.

The problem of parameter estimation (2) coincides, in our context, with the problem of maximum likelihood estimation (MLE). The computational challenge involved in using ERGMs is the intractable normalizing constant $k(\theta)$, that makes MLE computable only by Monte Carlo techniques[14]. Throughout the paper, we will be assuming that a MLE exists and that the model is non-degenerate[14,19]. It is known that this latter condition does not hold for all ERGM models (e.g., the edge-triangle model and other simple Markov random graph models[13,19,20] with phase transitions, where non-convergent estimations are common), but more robust specifications, specifically the "alternating" statistics used in this paper (see the beginning of Section Results for the exact ERGM specifications adopted), have been shown to be much better behaved in terms of stability of statistics across a wide range of parameter values[3].

Existing computational methods for MLE/MoM of ERGM parameters via equation (2) do not scale up easily to large data. Even though, to date, computational costs have constrained the scope of MLE, it remains widely adopted in numerous research settings, including the analysis of temporal networks[21,22]. Pseudo-likelihood and quasi-likelihood methods have been used when MLE cannot be computed, but it has been shown that these methods do not produce reliable results[20,23,24]. Estimation via conditional independence sampling methods, particularly snowball sampling, has been recently introduced to alleviate some of these issues of scale[25–27].

Estimates of the model parameters $\theta$ may be achieved using a number of computational approaches such as, for example, Markov chain Monte Carlo maximum likelihood estimation (MCMCMLE)[23,28–31], variants of stochastic approximation for the method of moments[18,22,32], and Bayesian estimation[33–35]. Of these different approaches, Bayesian estimation of ERGMs is the most computationally intensive due to the double intractability of the posterior distribution[34]. In other types of models, however, new methods based on variational Bayesian inference[36] can be relatively efficient, leveraging the sparse structure of large networks, e.g., graph partitioning (via blockmodeling analysis and/or analysis of modularity)[37,38]. These methods, however, have yet to be adopted for the estimation of ERGMs. In this paper, we derive an efficient Monte Carlo approach for MLE. MCMCMLE methods are somewhat faster than the MoM because they use importance sampling but are crucially dependent on the choice of the importance distribution. Stochastic approximation for the MoM is often more robust. For the network models we describe here, the existing MCMCMLE and MoM methods are similar in practical terms with respect to the network size that can be estimated. A considerable variety of stochastic approximation methods have been developed for different problems. In particular, efficient stochastic gradient methods[39] are often adopted for maximum likelihood estimation when many independent observations are available. The MoM[18] adopts the Robbins-Monro algorithm[40] and Polyak-Ruppert averaging[41] and is often used to solve (2) when the left side of (2) can be computed only by MCMC simulation[42]. In this paper, we compare the performance of our new approach with that of the MoM[18].

MCMC simulation and, in particular, the Metropolis-Hastings algorithm[43,44] may be used to generate a network $x$ for fixed values of the parameters $\theta$, which we denote by $x(\theta)$. Equation (2) formulates the inverse problem, i.e., to find the value of $\theta$ that, given $x_{obs}$, satisfy (2), which we denote by $\theta(x_{obs})$. Computing $\theta(x_{obs})$ is much more computationally expensive than computing $x(\theta)$. The largest networks for which these methods have been applied to find the MLE of ERGM parameters contain at most a few thousand nodes. In contrast to existing approaches for MLE, what we propose does not require the simulation of a large number of Markov chains until convergence. Rather, it relies on properties of the Markov chain at equilibrium. For this reason, we call the proposed approach equilibrium expectation (EE).

The remainder of this paper is organized as follows: in the next section we give a brief description of Markov chain Monte Carlo, and then propose a new method, Equilibrium Expectation, for MLE of ERGM parameters. Then in the Results section, we demonstrate the performance of the EE algorithm on some simulated and empirical networks, of sizes well beyond what is currently possible in practice.

## Markov Chain Monte Carlo

If the value of the normalizing constant $k(\boldsymbol{\theta})$ in equation (1) is not known, the values of probability distribution (1) cannot be computed. Markov chain Monte Carlo simulation is typically used to address this issue. MCMC simulation allows approximation of the target probability (1) and computation of expected properties $E_{\pi(\theta)}(z_A(x)) = \sum_A z_A(x) \pi(x, \boldsymbol{\theta})$ of the model. The Metropolis-Hastings algorithm uses a Markov process that asymptotically reaches a unique stationary distribution $\pi(\boldsymbol{\theta})$. Given that the system is in state $x$ the new state $x'$ is proposed with probability $q(x \rightarrow x')$. A Markov process is uniquely defined by its transition probabilities, $P(x \rightarrow x', \boldsymbol{\theta}) = q(x \rightarrow x') \alpha(x \rightarrow x', \boldsymbol{\theta})$, i.e., the probability of transitioning from any given state $x$ to any state $x'$. Detailed balance with respect to a given distribution $\pi(x, \boldsymbol{\theta})$, which in turn implies that $\pi(x, \boldsymbol{\theta})$ is a stationary distribution for the Markov chain, is satisfied if the acceptance probability of the new state is given by

$$\alpha(x \rightarrow x', \boldsymbol{\theta}) = \min\left\{1, \frac{q(x' \rightarrow x) \pi(x', \boldsymbol{\theta})}{q(x \rightarrow x') \pi(x, \boldsymbol{\theta})}\right\} \tag{3}$$

Formally, the Metropolis-Hastings algorithm may be written as Algorithm 1.

---

**Algorithm 1.** Metropolis-Hastings algorithm.

---

1:  Pick an initial state $x_0$ at random.
2:  **for** $t \leftarrow 0$ to $T$ **do**
3:      Propose move $x_t = x \rightarrow x'$ with probability $q(x \rightarrow x')$.
4:      Using (3), calculate Metropolis-Hastings acceptance probability $\alpha(x \rightarrow x', \boldsymbol{\theta})$.
5:      **if** $\text{Unif}(0,1) < \alpha(x \rightarrow x', \boldsymbol{\theta})$ **then**                 ▷ With probability $\alpha(x \rightarrow x', \boldsymbol{\theta})$,
6:          $x_{t+1} \leftarrow x'$                                                              ▷ the proposed move from $x_t = x \rightarrow x'$ is accepted.
7:      **else**
8:          $x_{t+1} \leftarrow x_t$
9:      **end if**
10: **end for**

---

If $t$ is larger than the burn-in time, $t_B$, of the Markov process, then $x_t$ values are drawn from $\pi(x, \boldsymbol{\theta})$, where $t_B$ is the time taken by the Markov chain to forget its initial state and reach the stationary regime. The algorithm generates $z_A(t) = z_A(x_t)$ sequences. From the Markov chain ergodic theorem it follows that, under regularity conditions, if the number of steps $T$ is large, then the expected values of the model statistics $E_{\pi(\theta)}(z_A(x))$ can be estimated by the average of $z_A(t)$ along the path of the Markov chain, $\overline{z_A(t > t_B)}$[44]. The Metropolis-Hastings algorithm provides a general framework that results in a large number of different MCMC samplers that differ in the proposal $q(x \rightarrow x')$. Verification of MCMC convergence is not a simple matter, and different methods have been suggested[45]. One heuristic rule for evaluating convergence of the Markov process $x_t$, in our ERGM context, is to monitor the $z_A(x_t)$ sequences.

Consider the expected change $E_{\pi(\theta)}(\Delta z_A)$ of $z_A$, $E_{\pi(\theta)}(\Delta z_A) = E_{\pi(\theta)}(z_A(x_{t+1}) - z_A(x_t))$. If equilibrium stationary distribution is reached, then statistics $z_A(x_t)$ converge and fluctuate around their expected values $E_{\pi(\theta)}(z_A)$, which do not depend on $t$, and hence

$$E_{\pi(\boldsymbol{\theta})}(\Delta z_A) = 0 \tag{4}$$

We can write $E_{\pi(\theta)}(\Delta z_A)$ as a function of the transition probabilities. Given $x_t = x$, the expected change along the path of the Markov chain in $z_A(x_{t+1}) - z_A(x_t)$ is obtained as

$$\Delta z_A(x, \boldsymbol{\theta}) = \sum_{x'} P(x \rightarrow x', \boldsymbol{\theta})(z_A(x') - z_A(x)) \tag{5}$$

and the expected value of $\Delta z_A(x, \boldsymbol{\theta})$ with respect to $\pi(x, \boldsymbol{\theta})$ is

$$E_{\pi(\boldsymbol{\theta})}(\Delta z_A) = \sum_{x,x'} \pi(x, \boldsymbol{\theta}) P(x \rightarrow x', \boldsymbol{\theta})(z_A(x') - z_A(x)) \tag{6}$$

From (4), it follows that after the MCMC burn-in, when equilibrium stationary distribution is reached, we have that for all the statistics $z_A(x)$,

$$\sum_{x,x'} \pi(x, \boldsymbol{\theta}) P(x \rightarrow x', \boldsymbol{\theta})(z_A(x') - z_A(x)) = 0 \tag{7}$$

## Estimation strategy: Equilibrium Expectation

In the context of ERGMs, MLE of parameters is obtained from equation (2). Existing computational methods for MLE, such as MCMCMLE, the MoM, or Bayesian estimation, use iterative algorithms that successively modify $\boldsymbol{\theta}$ until (2) is satisfied within given criteria for MCMC convergence. To this end, MCMC simulation is performed by drawing a large number of simulated networks for various values of $\boldsymbol{\theta}$, which we denote by $x_S(\boldsymbol{\theta})$. The simulated network $x_S$ is a network drawn from probability distribution $\pi(x, \boldsymbol{\theta})$.

Each time a simulated network $x_S(\theta)$ is drawn, the convergence criterion (7) should be satisfied, but this makes the standard estimation algorithms very computationally expensive. We suggest a much less computationally expensive approach for MLE.

We use equation (7) and rewrite it as:

$$E_{\pi(\theta)}(\Delta z_A(x, \theta)) = 0 \qquad (8)$$

Equation (8) reads as follows: if network $x$ is drawn from probability distribution $\pi(x, \theta)$, then the expected value of $\Delta z_A(x, \theta)$ is zero. Equation (8) is valid only at equilibrium, that is, when $\pi(x, \theta)$ is the limiting equilibrium distribution of the Markov chain that has transition probabilities $P(x \to x', \theta)$. $\Delta z_A(x, \theta)$ may be computed via Monte Carlo integration[42,46,47], as suggested in Section Estimation algorithm. The expectation with respect to $\pi(x, \theta)$ could be computed if a large Monte Carlo sample of networks independent and identically distributed (i.i.d.) from $\pi(x, \theta)$ were available i.e., $x_{S_1}, x_{S_2}, \cdots, x_{S_n}$. The existence of this large i.i.d. sample of networks is assumed here for expository purposes in deriving an estimation strategy, and later we will remove this assumption to develop an estimation method which can be applied to observed empirical networks. Making this assumption, the LHS of (8) may be computed by Monte Carlo integration as $E_{\pi(\theta)}(\Delta z_A(x\theta)) = \frac{1}{n}\sum_i \Delta z_A(x_{S_i}, \theta)$, and (8) may be approximated by

$$\frac{1}{n}\sum_i \Delta z_A(x_{S_i}, \theta) = 0 \qquad (9)$$

If we have a large sample of networks i.i.d. from $\pi(x, \theta)$, then we can efficiently compute the LHS of (9) and solve system of equations (9) with respect to $\theta$. Thus, we can estimate $\theta$. When MCMC simulation is performed, the number of steps should be larger than the burn-in time, which may be large. In contrast, $\Delta z_A(x, \theta)$ in (9) may be computed by Monte Carlo integration, there is no burn-in, and the number of steps may be small.

Equation (2) may be written as $f_A(x_{obs}, \theta) = 0$, where $f_A(x_{obs}, \theta) = E_{\pi(\theta)}(z_A(x) - z_A(x_{obs}))$; hence, the true parameter values, $\theta^*$, may be estimated from

$$\frac{1}{n}\sum_i f_A(x_{S_i}, \theta^*) = 0 \qquad (10)$$

Typically, to estimate $\theta^*$, the estimation of each observed network is performed, and the resulting estimates $\theta(x_{S_i})$ are averaged over the observations as follows: $\hat{\theta}^* = \frac{1}{n}\sum_i \theta(x_{S_i})$. If the network $x_S$ is very large, then $\theta(x_{S_i})$ is the desired estimate of the true $\theta^*$. Thus, if the network $x_S$ is very large, then the true $\theta^*$ may be estimated from $f_A(x_{obs}, \theta) = 0$, and the summation in equation (10) may be dropped. In statistics and statistical physics, this property is called the ergodicity of systems: the time averaging is equivalent to the ensemble averaging. Here, time averaging is the average over networks generated by the Markov process, and ensemble averaging is the average over the space of all the system's states, which grows with the network size. If the system is ergodic, then this property may also be applied to equation (9) i.e., if the network $x_s$ is very large, then the summation in equation (9) may be dropped. Thus, the true $\theta^*$ value may be found from the following equilibrium expectation condition: for all $A$,

$$\Delta z_A(x_S, \theta^{EE}) = 0, \qquad (11)$$

where $\Delta z_A(x_S, \theta)$ is given by (5). We can find the value of $\theta^{EE}$ that, given $x_S$, satisfies (11); this value is denoted as $\theta^{EE}(x_S)$. If the network $x_S$ is very large, then $\theta^{EE}(x_S)$ is the desired estimate of $\theta^*$. Otherwise, a large sample of networks is required as described above, and the true $\theta^*$ value may be found from (9).

The estimation method described above may be theoretically interesting, but is of limited value for the estimation of real-world data, for which we have only a single observed network $x_{obs}$ drawn from some unknown probability distribution, rather than a large i.i.d. Monte Carlo sample of networks drawn from a known probability distribution, as we assumed earlier.

It is not, however, lacking in usefulness when applied to such an observed network. In fact, it is actually Contrastive Divergence[48] (CD) as applied to ERGM parameter estimation[31,49–51]. The CD algorithm, rather than running an MCMC simulation to convergence, instead starts from the observed data and makes only some number $k$ of MCMC updates (the CD algorithm with $k$ updates is called CD-$k$). CD-1 has been shown to be equivalent to maximum pseudo-likelihood under certain conditions[52] and it has also been shown that CD-$k$ forms a series of increasingly close approximations to the MLE as $k \to \infty$[49].

To see that the estimation strategy we have just described is equivalent (but independently developed) to CD-1, it may be useful to consider the implementation described by Krivitsky[31] of using CD to find the initial estimates to seed MCMCMLE. In this implementation, an ERGM MCMCMLE implementation is simply converted to CD-$k$ by modifying the MCMC sampler to start from $x_{obs}$ and reverting the chain to $x_{obs}$ every $k$ steps. And so if $k = 1$ then the network is not actually modified and the CD-1 estimate of $\theta$ is just the solution of (10). Details of the CD-1 algorithm as applied in the work described here are given in the Supplementary Information.

As previously noted, maximum pseudo-likelihood (and hence CD-1) does not necessarily produce reliable results for ERGM estimation. However they (and CD-$k$ for $k > 1$) are useful for finding initial $\theta$ values to seed methods such as MCMCMLE that are sensitive to initialization conditions[31,53]. In this paper we indeed use CD-1 to find initial parameter estimates.

We will now describe how this strategy can be modified, for more realistic applications, having an observed network drawn from some unknown probability distribution. Let $\pi^*(x)$ be the unknown probability distribution from which the real-world data $x_{obs}$ are drawn. $\pi(x, \boldsymbol{\theta})$ is then the probability distribution corresponding to the model specification we have chosen (in this paper the ERGM model for some specified choices of network statistics). We assume this model to be appropriate for the observed data – we do not discuss model selection here – and, as discussed in the Introduction, we also assume the MLE exists and the model is not degenerate. Formally, we can write that for any $\boldsymbol{\theta}$, $\pi^*(x) \neq \pi(x, \boldsymbol{\theta})$. Equation (7) can be applied only if the equilibrium stationary distribution of the Markov chain follows a statistical model $\pi(x, \boldsymbol{\theta})$. If $\pi^*(x) \neq \pi(x, \boldsymbol{\theta})$, then the conditions of Equation (7) are not satisfied, and it cannot be applied. In other words, if observed networks are not drawn from $\pi(x, \boldsymbol{\theta})$, then the LHS of equation (7) is not equal to the LHS of equation (9). In this case, we cannot compute the LHS of (7), and hence we cannot find a value of $\boldsymbol{\theta}$ such that (7) is satisfied. However, if a solution to (2) exists, then it is possible to find $x_S$ drawn from $\pi(x, \boldsymbol{\theta})$ such that the following condition is satisfied:

$$z_A(x_S) = z_A(x_{obs}). \tag{12}$$

For network $x_S$, drawn from probability distribution $\pi(x, \boldsymbol{\theta})$, (11) may be used as described above. Using (10) and (12), the solution of (2) may be obtained for an observed network drawn from an unknown probability distribution. The network drawn from $\pi(x, \boldsymbol{\theta})$ may be obtained by an MCMC simulation. In the estimation strategy just described, the parameters $\boldsymbol{\theta}$ are iteratively adjusted until (11) is satisfied. To estimate parameters for networks drawn from an unknown distribution $\pi^*(x)$, both $\boldsymbol{\theta}$ and $x$ are modified: $\boldsymbol{\theta}$ is modified in the same way as before, and $x$ is also modified (equilibrated) so that, after some burn-in, $x$ will be drawn from the probability distribution $\pi(x, \boldsymbol{\theta})$.

Indeed, starting from the previously described method to find the solution of (11), we can derive the EE algorithm (described in the following section) for the solution of (11), (12). The EE algorithm for the estimation of an observed network $x_{obs}$ performs Metropolis-Hastings moves, that is, it actually makes changes to a network for accepted proposals, a step which was not necessary for solving (11). The EE algorithm requires MCMC simulation, but – in contrast to MCMCMLE, the MoM or Bayesian estimation – the EE algorithm does not draw many simulated networks for various $\boldsymbol{\theta}$ values and, therefore, is considerably faster.

## Estimation algorithm

The EE algorithm is described in Algorithm 2. To provide some intuition, the description is maintained at a rather general level here. A more detailed and formal description of the EE algorithm is included in the Supplementary Information. Let $m = 1000$ and $M$ indicate the number of steps of the EE algorithm. $K_A$ is a positive constant, the values of which could be for example $K_A \approx 10^{-4} \cdot (\partial \Delta z_A(x_{obs}, \boldsymbol{\theta})/\partial \theta_A)^{-2}$; a better choice is suggested in the Supplementary Information. The final parameter estimate from the CD-1 algorithm is used as the starting point for the EE algorithm, but the efficiency of the EE algorithm is more sensitive to $K_A$ values than to the initial parameter estimates $\boldsymbol{\theta}_0$.

Steps 4–5 of the EE algorithm are identical to steps 3–4 of the Metropolis-Hastings algorithm presented in Section Chain Monte Carlo: the move $x \rightarrow x'$ is proposed, and the acceptance probability (3) is calculated. If the move to $x'$ is accepted, then the change in the sufficient statistics $z_A(x') - z_A(x)$ is computed.

Step 13 modifies $\boldsymbol{\theta}$ iteratively until (11) is satisfied. This parameter update depends on the property of the exponential family that $E_{\pi(\boldsymbol{\theta})}(z_A(x))$ is a monotonically increasing function of $\theta_A$, which was mentioned in the Introduction. Because of this, for each configuration $A$ we want to increase $\theta_A$ when $dz_A$ is negative and decrease it when $dz_A$ is positive, in order to find $\theta_A$ such that $dz_A$ is zero. This is achieved by adding $-K_A \cdot \text{sign}(dz_A)dz_A^2$ to the current $\theta_A$ value (recall that $K_A$ is positive).

Different root-finding algorithms may be applied to solve the system of equation (11) when $\Delta z_A(x, \boldsymbol{\theta})$ is available. For example, bisection or quasi-Newton methods[46,54] may be applied if $\Delta z_A(x, \boldsymbol{\theta})$ were a continuous functions of $\boldsymbol{\theta}$. Efficient root finding methods use derivative information. The precise derivative information is either not available or very computationally expensive, but it may be shown (see Supplementary Information) that, for any $x$,

**Algorithm 2.** EE Algorithm.

```
 1: Initialization: x ← x_obs; dz ← 0
 2: for t ← 0 to M − 1 do
 3:     for k ← 1 to m do
 4:         Propose move x_t = x → x' with probability q(x → x')
 5:         Using (3), calculate Metropolis-Hastings acceptance probability α(x → x', θ_t)
 6:         if Unif(0, 1) < α(x → x', θ_t) then                          ▷ With probability α(x → x', θ_t)
 7:             dz_A ← dz_A + z_A(x') − z_A(x) for all configurations A
 8:             x_{t+1} ← x'                                              ▷ the proposed move from x_t = x → x' is accepted.
 9:         else
10:             x_{t+1} ← x_t
11:         end if
12:     end for
13:     Update of parameters θ_{t+1,A} ← θ_{t,A} − K_A · sign(dz_A)dz_A^2 for all configurations A
14:     Save sequences dz_A(t + 1) ← dz_A(t) for all configurations A
15: end for
```

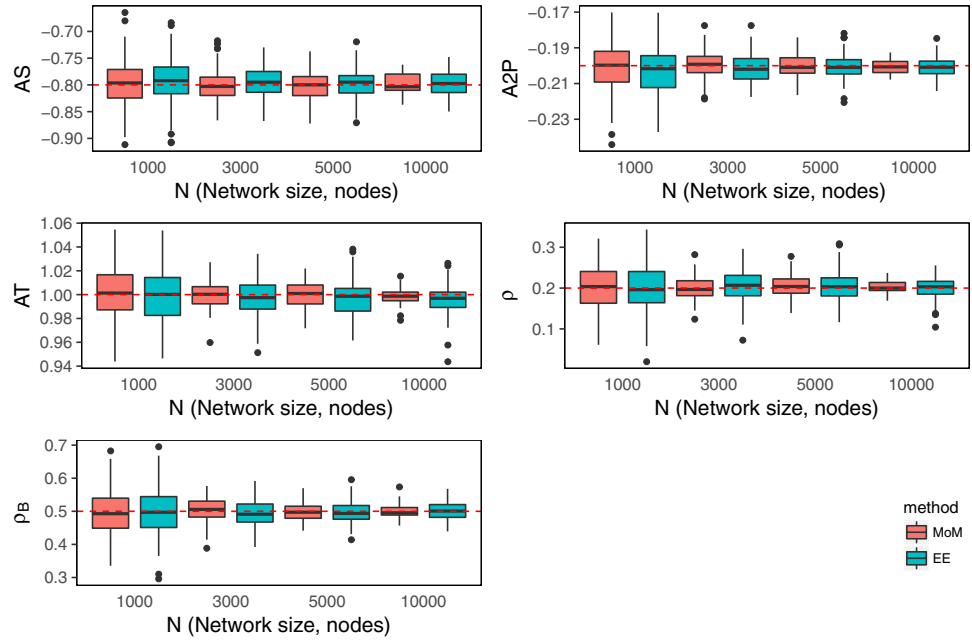

**Figure 1.** Estimation of simulated networks by suggested EE algorithm and MoM[18]. Horizontal lines show the true $\theta_A^*$ value and the estimates from each of the two methods for each of the four network sizes are shown as boxplots (generated with ggplot2[65]). Each boxplot represents estimates for 120 networks, except that for MoM, when $N = 5000$ only 118 estimations converged and when $N = 10{,}000$ only 49 of the 120 estimations converged.

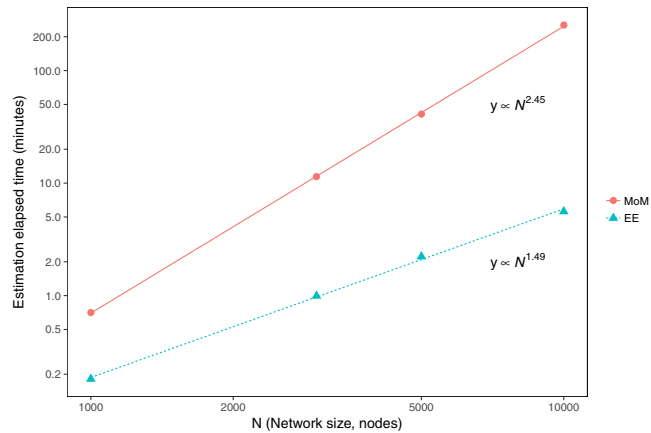

**Figure 2.** Estimation times (observed values for different network sizes are reported as circles and triangles) for the EE algorithm and Method of Moments (MoM) with fitted lines. Both axes are on a log scale.

$$\partial \Delta z_A(x, \boldsymbol{\theta})/\partial \theta_A \geq 0. \tag{13}$$

It is trivial to find roots of continuous monotonic functions, and step 13 implements one possibility. A more robust approach to solving $\Delta z_A(x, \boldsymbol{\theta}) = 0$ would be to use stochastic approximation methods, which are typically applied to solve (2).

Since $x$ is not constant, $z_A(x)$ is also not constant. Recently, an auxiliary parameter MCMC method was proposed to perform MCMC simulation, constraining the value of one of the network statistics. Only the network statistic $z_L(x)$, which is the edge count, was constrained to the observed value, and this was achieved by using a special proposal $q(x \rightarrow x')$: the IFD sampler[55]. The corresponding parameter $\theta_L$ was adapted[55] in the same way as done in step 13 of the EE algorithm. In the present paper, all the network statistics $z_A(x)$ are constrained and, crucially, without the need to develop a special proposal distribution. Indeed, they may be constrained by exploiting (13), i.e., the monotonic dependence of $\Delta z_A(x, \boldsymbol{\theta})$ and $E_{\pi(\boldsymbol{\theta})}(z_A(x))$ on $\theta_A$, and by properly modifying the $\theta_A$ values. We need an algorithm that converges to the values of $\boldsymbol{\theta}$ and $x$ that satisfy (11) and (12) for all the network statistics $z_A(x)$. This is achieved by accumulating the accepted change statistics (line 7). As a result, $dz_A = dz_A(t) + \Delta z_A(x, \boldsymbol{\theta})$ and $dz_A = z_A(x) - z_A(x_{obs})$. If equilibrium is reached, then $x = x_S$ is drawn from $\pi(x, \boldsymbol{\theta})$ and hence $\Delta z_A(x, \boldsymbol{\theta}) = 0$ and (11) is satisfied. In addition, if $dz_A = 0$, then $z_A(x) = z_A(x_{obs})$ and (12) is satisfied. It is shown in Section Estimation strategy: Equilibrium Expectation that the solution of (11) and (12) gives the solution of (2).

| AT | Mismatch E class | Mismatch kinase-phosphorylated | Edge (L) | Isolates | AS | Activity plant specific | Interaction plant specific |
|---|---|---|---|---|---|---|---|
| 1.276 | 1.304 | 0.192 | −14.940 | −7.116 | 2.320 | −0.104 | 0.456 |
| (1.24, 1.31) | (0.77, 1.83) | (0.08, 0.30) | (−14.97, −14.92) | (−7.59, −6.64) | (2.23, 2.41) | (−0.15, −0.06) | (0.21, 0.70) |

**Table 1.** Parameter estimates with 95% confidence interval (see Supplementary Information) for the *Arabidopsis thaliana* PPI network, estimated using the EE algorithm with the IFD sampler. Estimation of this 2,160 nodes network took only 3 minutes on the Lenovo NeXtScale x86 system at Melbourne Bioinformatics.

The EE algorithm produces $\theta_A(t)$ sequences. The number of steps $M$ should be large enough for $\theta_A(t)$ to converge. After $\theta_A(t)$ reaches convergence, its values fluctuate around some constant values $\theta_A$, which we estimate as $\theta_A = \overline{\theta_A(t > t_B)}$. In addition to $\theta_A(t)$, the sequences $dz_A(t) = z_A(t) - z_A(x_{obs})$ are generated by the EE algorithm. It is convenient to plot both $z_A - z_A(x_{obs})$ and $\theta_A(t)$ sequences to visually check for convergence of the algorithm. We use the *t*-ratio test to check if (12) is satisfied:

$$\tau_A = \frac{\overline{z_A(t > t_B)} - z_A(x_{obs})}{SD(z_A(t > t_B))},\qquad(14)$$

where $t_B$ is the burn-in time, starting from which convergence is observed from the plots of $\theta_A(t)$ and $z_A(t) - z_A(x_{obs})$, and $\overline{z_A(t > t_B)}$ denotes the network statistic averaged over $t > t_B$, while $SD(\cdot)$ is the standard deviation operator. We propose an appropriate indication of good convergence so that the suggested algorithm gives a converged solution to (2) if, for all $A$, (i) $|\tau_A| < 0.1$ and (ii) $\theta_A$ converges.

Note that if $K_A = 0$, then the EE algorithm does not differ from the Metropolis-Hastings algorithm. MCMC estimation algorithms make use of MCMC simulation; hence, the efficiency of MCMC estimation algorithms depends on the efficiency of MCMC samplers. In turn, the efficiency of the Metropolis-Hastings algorithm depends on the proposal $q(x \rightarrow x')$. The same holds also for the EE algorithm: the algorithm may be used with different proposals $q(x \rightarrow x')$, and its efficiency depends on a good choice of the proposal distribution. Proposals for ERGMs are reviewed elsewhere[55]. Both the updating step of the MoM[18] and that of the EE algorithm depend on $z_A(x) - z_A(x_{obs})$. However, in the case of the MoM, $x$ is the equilibrium network configuration, which can be obtained by, e.g., the Metropolis-Hastings algorithm if the MCMC simulation time is larger than the burn-in time of the Markov process. In the EE algorithm, $x$ is a current non-equilibrium state. The MoM and MCMCMLE[28] require many converged outputs of the Metropolis-Hastings algorithm, while the EE algorithm does not need such outputs. Instead, the EE algorithm generates one converged output.

## Results

First, we test the EE algorithm by computing the bias and the standard deviation of the estimates that this algorithm produces. Simulated networks of various sizes $N$ (here, $N$ is the number of nodes in a network) are generated by the Metropolis-Hastings algorithm (as implemented in the PNet[56] program). The ERGM model is defined by the AS, AT, A2P, $\rho$ and $\rho_B$ network statistics. The network statistics we use in this paper were suggested by Snijders et al.[3,20] and are detailed in the Supplementary Information. The true values of the corresponding model parameters $\theta_A^*$ are represented by the horizontal lines in Fig. 1, with the estimated values of the parameters shown as boxplots, for both the EE algorithm and a variant of the stochastic approximation algorithm (MoM)[18] for MLE, widely used for ERGM estimation. It is clear that both EE and the MoM give accurate estimates of the true $\theta_A^*$ values. Both the variance and the bias $|\theta_A - \theta_A^*|$ of the estimates obtained using different methods have similar values. In Fig. 2, we report the computational times needed to obtain the corresponding estimates. As described in the Methods section, steps 4–6 of the EE algorithm are equivalent to steps 3–5 of the Metropolis-Hastings algorithm, used in the MoM. In our implementation of the MoM and EE algorithms, these steps are carried out in exactly the same way. This allows the evaluation of the efficiency of the MoM and EE algorithms by comparing the computational times for the corresponding estimations. Both the MoM and EE algorithms are used with the Basic sampler[55], and estimation is performed on a Cray XC50 machine available at the Swiss National Supercomputing Centre (CSCS). Figure 2 shows that for small networks ($N = 1,000$), the efficiencies of the EE and MoM algorithms are not very different, while for larger networks ($N = 10,000$), the EE algorithm is 2 orders of magnitude faster than the MoM. These results suggest that the EE algorithm may be used for the analysis of large datasets.

For a more complete comparison of the MoM and EE, we now compare their respective performance on empirical datasets. Using the MoM, it is possible to estimate simulated networks with 10,000 nodes within several hours, but the estimation of empirical networks takes longer to converge. Despite several attempts to improve the efficiency of the computational methods for MLE, the existing algorithms are still too computationally expensive for empirical networks of this or larger size. Using the more efficient IFD sampler[55], we can estimate six biological networks via both EE and the MoM. The estimation results are reported in the Supplementary Information. The results reported in Supplementary Tables S2 and S3 show that the MoM and EE produce equivalent estimates, but EE may be two orders of magnitude faster than the MoM, even for relatively small empirical networks with 1,781 and 5,038 nodes. Although we cannot estimate larger networks with the MoM, the results reported in Fig. 2 suggest that for larger networks, EE may be orders of magnitude faster.

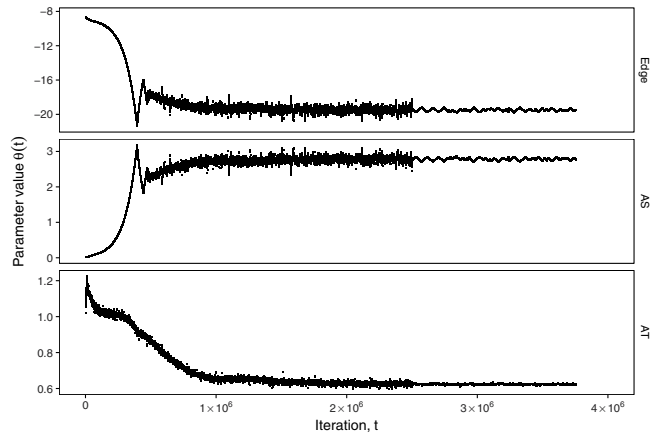

**Figure 3.** Results of estimation of ERGM parameters for *Livemocha* networks with 104,103 nodes and 2,193,083 ties using the EE algorithm. The starting point is the result of the CD-1 algorithm. Producing these results took 12 hours on one core of the Intel E5-2650 machines available at https://intranet.ics.usi.ch/HPC.

For an illustrative example of a biological network, we study an *Arabidopsis thaliana* protein-protein interaction (PPI) network[57,58], including some protein attributes on the nodes. In this network, nodes represent proteins, and (undirected) edges represent literature-curated binary interactions between proteins. *Arabidopsis thaliana* proteins are annotated with various properties, as described in the Supporting Online Material of Arabidopsis Interactome Mapping Consortium[58]. We make use of the following protein attributes. We consider one binary "plant-specific" node attribute (genes defined as plant-specific, absent from other eukaryotic lineages[58]) and define activity $\rho$ and interaction $\rho_B$ network statistics on this attribute. We consider 2 categorical attributes "E class" and "kinase-phosphorylated", derived from the Supporting Online Material of Arabidopsis Interactome Mapping Consortium[58] as described in the Supplementary Information, and define mismatch network statistics on these attributes. The estimated parameters of the resulting model are given in Table 1. The significant positive AT parameter indicates that triangle motifs are significantly over-represented. The positive plant-specific interaction parameter shows that plant-specific proteins interact preferentially with each other (with the negative activity effect showing that, relative to other proteins, they are less likely to interact at all). The positive mismatch kinase-phosphorylated effect shows the propensity of kinases to interact with phosphorylated proteins, as we would expect given the action of kinases to phosphorylate proteins. The positive mismatch E class shows the propensity of proteins to interact if they belong to different E classes[59].

We also demonstrate the performance of the algorithm in a case study involving the analysis of a larger empirical social network. More specifically, we study the *Livemocha* network of friendship relations among members of an online language learning community available from the repository at http://konect.uni-koblenz.de [60,61]. Founded in 2007, *Livemocha* was an online community of peer-to-peer language learning that provided language instruction to approximately 12 million users from more than 190 countries. *Livemocha* closed in 2016 after being included by the *Times* magazine in its list of 50 best websites in 2010. The *Livemocha* dataset that we analyze consists of an undirected network with 104,103 nodes and 2,193,083 ties. We estimate AS, AT and edge parameters using the EE algorithm and IFD sampler. The convergence criteria $|\tau_A| < 0.1$ is satisfied, and the values of estimated parameters $\theta_A(t)$ are presented in Fig. 3 as a function of the algorithm step $t$. One can see that $\theta_A(t)$ converge. The $\theta_A(t=0)$ values are obtained using the CD-1 algorithm, and the smaller $K_A$ values of the EE algorithm are used for $t > 2.5 \times 10^6$ (the computational details are given in the Supplementary Information). The values of the estimated parameters are significantly different from zero. The $\theta_A(t)$ values are averaged over the last $5 \times 10^5$ steps, and the following estimates of the model parameters are obtained: $\theta_L = -19.321$, $\theta_{AS} = 2.7355$, and $\theta_{AT} = 0.6453$. A different convergence test was suggested by Snijders for the MoM algorithm: estimated parameter values may be used to estimate $E_{\pi(\theta)}(z_A(x))$ in (2) via the Metropolis-Hastings algorithm, and the t-statistics[18] may be computed. We checked that this convergence test is also satisfied: for all $A$, the absolute values of the t-statistics are less than 0.3. However, the convergence test, as suggested by Snijders, is computationally expensive. When the EE algorithm is applied, it is more convenient to check the values of $\tau_A$ and to visualize $\theta_A(t)$ dependencies.

### Discussion

We propose a Monte Carlo Markov chain based approach to the maximum likelihood estimation of parameters of statistical models with intractable normalizing constants belonging to the linear exponential family. The algorithm we propose is similar to the Metropolis-Hastings algorithm, but allows Monte Carlo simulation to be performed while constraining the values of model statistics so that the equilibrium expectation condition (11) is satisfied. In contrast to the widely adopted Metropolis-Hastings algorithm, which is guaranteed to converge within the limit of infinite simulation time, the EE algorithm provides no such guarantee. Further work is required to establish the general convergence conditions of the algorithm. This work is currently in progress.

We demonstrate the merits of the algorithm by estimating the parameters of ERGMs on large network data sets. ERGMs are popular statistical models for the analysis of single-observation network data. ERGMs may be used to determine motif significance, and are widely used to connect motifs to structural features in social and other kinds of networks. We compute the bias of the estimates produced by the EE algorithm and show that its value is close to the bias of a commonly adopted computational method for MLE. However, EE is faster, and the estimation time increases with the number of network nodes as $N^{1.5}$. These results suggest that the EE algorithm may be relied upon to support statistical inference on very large complex systems with network-like local dependencies. We then study several biological networks and a large social network. We show that the triangle motif is significantly over-represented in all the networks studied. We show that accurate maximum likelihood estimates of ERGM parameters may be obtained for a large empirical network with 104,103 nodes and 2,193,083 ties. The smallest network for which we applied the EE algorithm is a network with 418 nodes (see Supplementary Information).

The algorithm is inappropriate for the curved exponential family[30] when the number of parameters to be estimated differs from the number of statistics. Currently, we are applying the EE algorithm to study directed networks for which different network statistics need to be computed[3]. Future research directions involve an attempt to couple the EE algorithm with the expectation-maximization algorithm[62] for incomplete data, and an extension to bipartite networks. In addition, we believe the computational approach we have proposed in this paper may be applicable to a wider range of statistical models for data characterized by complex network-like dependencies.

**Data availability.** The datasets analyzed in the current study are available in the following repositories: Livemocha http://konect.uni-koblenz.de/networks/livemocha, http://socialcomputing.asu.edu/datasets/Livemocha; *A. thaliana* PPI: http://interactome.dfci.harvard.edu/A_thaliana; Yeast PPI: igraph[63] Nexus repository (as of March 2017, the igraph Nexus repository is no longer available; we downloaded this data set on 10 November 2016, and it is available upon request); Human PPI: https://icon.colorado.edu/#!/networks, http://interactome.dfci.harvard.edu/H_sapiens; *C. elegans* PPI: (downloaded from the web address specified in Huang *et al.*[64] on 2 March 2017, which is no longer available, data available on request); *E. coli* regulatory: http://www.statnet.org/index.shtml; *Drosophila* optic medulla: https://icon.colorado.edu/#!/networks, http://openconnecto.me/graph-services/download/.

**Code availability.** The EE algorithm is implemented in the Estimnet program, available from the authors upon request and from the webpage http://www.estimnet.org. The algorithm may be accessed from the webpage https://github.com/Byshkin/EquilibriumExpectation.

## References

 1. Borgatti, S. P., Mehra, A., Brass, D. J. & Labianca, G. Network analysis in the social sciences. *Science* **323**, 892–895 (2009).
 2. Butts, C. T. Revisiting the foundations of network analysis. *Science* **325**, 414–416 (2009).
 3. Snijders, T. A. B., Pattison, P. E., Robins, G. L. & Handcock, M. S. New specifications for exponential random graph models. *Sociol. Methodol.* **36**, 99–153 (2006).
 4. Lusher, D., Koskinen, J. & Robins, G. *Exponential random graph models for social networks: Theory, methods, and applications* (Cambridge University Press, 2013).
 5. Saul, Z. M. & Filkov, V. Exploring biological network structure using exponential random graph models. *Bioinformatics* **23**, 2604–2611 (2007).
 6. Barndorff-Nielsen, O. *Information and exponential families in statistical theory* (John Wiley & Sons, 2014).
 7. Geman, S. & Geman, D. Stochastic relaxation, Gibbs distributions, and the Bayesian restoration of images. In *Readings in Computer Vision*, 564–584 (Elsevier, 1987).
 8. Milo, R. *et al.* Network motifs: simple building blocks of complex networks. *Science* **298**, 824–827 (2002).
 9. Shen-Orr, S. S., Milo, R., Mangan, S. & Alon, U. Network motifs in the transcriptional regulation network of *Escherichia coli*. *Nat. Genet.* **31**, 64–68 (2002).
10. Artzy-Randrup, Y., Fleishman, S. J., Ben-Tal, N. & Stone, L. Comment on "Network motifs: simple building blocks of complex networks" and "Superfamilies of evolved and designed networks". *Science* **305**, 1107 (2004).
11. Ciriello, G. & Guerra, C. A review on models and algorithms for motif discovery in protein–protein interaction networks. *Brief. Funct. Genomic. Proteomic.* **7**, 147–156 (2008).
12. Kovanen, L., Kaski, K., Kertész, J. & Saramäki, J. Temporal motifs reveal homophily, gender-specific patterns, and group talk in call sequences. *Proc. Natl. Acad. Sci. USA* **110**, 18070–18075 (2013).
13. Frank, O. & Strauss, D. Markov graphs. *J. Am. Stat. Assoc.* **81**, 832–842 (1986).
14. Hunter, D. R., Krivitsky, P. N. & Schweinberger, M. Computational statistical methods for social network models. *J. Comput. Graph. Stat.* **21**, 856–882 (2012).
15. Newman, M. E., Watts, D. J. & Strogatz, S. H. Random graph models of social networks. *Proc. Natl. Acad. Sci. USA* **99**, 2566–2572 (2002).
16. Newman, M. E. & Clauset, A. Structure and inference in annotated networks. *Nat. Commun.* **7**, 11863 (2016).
17. Pallotti, F., Lomi, A. & Mascia, D. From network ties to network structures: Exponential random graph models of interorganizational relations. *Qual. Quant.* **47**, 1665–1685 (2013).
18. Snijders, T. A. B. Markov chain Monte Carlo estimation of exponential random graph models. *J. Soc. Struct.* **3**, 1–40 (2002).
19. Handcock, M. S. Statistical models for social networks: Inference and degeneracy. In *Dynamic Social Network Modeling and Analysis: Workshop Summary and Papers*, 229–240 (National Academies Press, 2003).
20. Robins, G., Snijders, T. A. B., Wang, P., Handcock, M. & Pattison, P. Recent developments in exponential random graph (p*) models for social networks. *Soc. Networks* **29**, 192–215 (2007).
21. Snijders, T. A. B., Koskinen, J. & Schweinberger, M. Maximum likelihood estimation for social network dynamics. *Ann. Appl. Stat.* **4**, 567–588 (2010).
22. Snijders, T. A. B. The statistical evaluation of social network dynamics. *Sociol. Methodol.* **31**, 361–395 (2001).
23. Hummel, R. M., Hunter, D. R. & Handcock, M. S. Improving simulation-based algorithms for fitting ERGMs. *J. Comput. Graph. Stat.* **21**, 920–939 (2012).
24. van Duijn, M. A., Gile, K. J. & Handcock, M. S. A framework for the comparison of maximum pseudo-likelihood and maximum likelihood estimation of exponential family random graph models. *Soc. Networks* **31**, 52–62 (2009).
25. Pattison, P. E., Robins, G. L., Snijders, T. A. B. & Wang, P. Conditional estimation of exponential random graph models from snowball sampling designs. *J. Math. Psychol.* **57**, 284–296 (2013).
26. Stivala, A. D., Koskinen, J. H., Rolls, D. A., Wang, P. & Robins, G. L. Snowball sampling for estimating exponential random graph models for large networks. *Soc. Networks* **47**, 167–188 (2016).

27. Thiemichen, S. & Kauermann, G. Stable exponential random graph models with non-parametric components for large dense networks. *Soc. Networks* **49**, 67–80 (2017).
28. Geyer, C. J. & Thompson, E. A. Constrained Monte Carlo maximum likelihood for dependent data. *J. Roy. Stat. Soc. B Met.* **54**, 657–699 (1992).
29. Geyer, C. J. Markov chain Monte Carlo maximum likelihood. In Keramides, E. M. (ed.) *Computing Science and Statistics: Proceedings of the 23rd Symposium on the Interface*, 156–163 (Interface Foundation of North America, 1991).
30. Hunter, D. R. & Handcock, M. S. Inference in curved exponential family models for networks. *J. Comput. Graph. Stat.* **15**, 565–583 (2006).
31. Krivitsky, P. N. Using contrastive divergence to seed Monte Carlo MLE for exponential-family random graph models. *Comput. Stat. Data Anal.* **107**, 149–161 (2017).
32. Okabayashi, S. *et al.* Long range search for maximum likelihood in exponential families. *Electron. J. Stat.* **6**, 123–147 (2012).
33. Lehmann, E. L. & Casella, G. *Theory of point estimation* (Springer Science & Business Media, 2006).
34. Caimo, A. & Friel, N. Bayesian inference for exponential random graph models. *Soc. Networks* **33**, 41–55 (2011).
35. Liang, F., Jin, I. H., Song, Q. & Liu, J. S. An adaptive exchange algorithm for sampling from distributions with intractable normalizing constants. *J. Am. Stat. Assoc.* **111**, 377–393 (2016).
36. Blei, D. M., Kucukelbir, A. & McAuliffe, J. D. Variational inference: A review for statisticians. *J. Am. Stat. Assoc.* **112**, 859–877 (2017).
37. Zhang, P. & Moore, C. Scalable detection of statistically significant communities and hierarchies, using message passing for modularity. *Proc. Natl. Acad. Sci. USA* **111**, 18144–18149 (2014).
38. Zhang, P., Krzakala, F., Reichardt, J. & Zdeborová, L. Comparative study for inference of hidden classes in stochastic block models. *J. Stat. Mech. Theory Exp.* **2012**, P12021 (2012).
39. Roux, N. L., Schmidt, M. & Bach, F. R. A stochastic gradient method with an exponential convergence rate for finite training sets. *Adv. Neural Inf. Process. Syst.* **25**, 2663–2671 (2012).
40. Robbins, H. & Monro, S. A stochastic approximation method. *Ann. Math. Stat.* **40**, 400–407 (1951).
41. Polyak, B. T. & Juditsky, A. B. Acceleration of stochastic approximation by averaging. *SIAM J. Contr. Optim.* **30**, 838–855 (1992).
42. Christian, P. R. & Casella, G. *Monte Carlo statistical methods* (Springer, 1999).
43. Metropolis, N., Rosenbluth, A. W., Rosenbluth, M. N., Teller, A. H. & Teller, E. Equation of state calculations by fast computing machines. *J. Chem. Phys.* **21**, 1087–1092 (1953).
44. Hastings, W. K. Monte Carlo sampling methods using Markov chains and their applications. *Biometrika* **57**, 97–109 (1970).
45. Cowles, M. K. & Carlin, B. P. Markov chain Monte Carlo convergence diagnostics: a comparative review. *J. Am. Stat. Assoc.* **91**, 883–904 (1996).
46. Miranda, M. J. & Fackler, P. L. *Applied computational economics and finance* (MIT press, 2004).
47. Newman, M. & Barkema, G. *Monte Carlo Methods in Statistical Physics chapter 1–4* (Oxford University Press: New York, USA, 1999).
48. Hinton, G. E. Training products of experts by minimizing contrastive divergence. *Neural Comput.* **14**, 1771–1800 (2002).
49. Asuncion, A., Liu, Q., Ihler, A. & Smyth, P. Learning with blocks: Composite likelihood and contrastive divergence. In *Proceedings of the Thirteenth International Conference on Artificial Intelligence and Statistics*, 33–40 (2010).
50. Hummel, R. M. *Improving estimation for exponential-family random graph models*. Ph.D. thesis, The Pennsylvania State University, https://etda.libraries.psu.edu/catalog/11493 (2010).
51. Fellows, I. E. Why (and when and how) contrastive divergence works. *arXiv preprint arXiv:1405.0602* (2014).
52. Hyvärinen, A. Consistency of pseudolikelihood estimation of fully visible Boltzmann machines. *Neural Comput.* **18**, 2283–2292 (2006).
53. Carreira-Perpiñan, M. A. & Hinton, G. E. On contrastive divergence learning. In *Proceedings of the Tenth International Workshop on Artificial Intelligence and Statistics*, 33–40 (2005).
54. Ortega, J. M. & Rheinboldt, W. C. *Iterative solution of nonlinear equations in several variables* (SIAM, 1970).
55. Byshkin, M. *et al.* Auxiliary parameter MCMC for exponential random graph models. *J. Stat. Phys.* **165**, 740–754 (2016).
56. Wang, P., Robins, G. & Pattison, P. *PNet: program for the estimation and simulation of p\* exponential random graph models*. Department of Psychology, The University of Melbourne (2009).
57. Swarbreck, D. *et al.* The *Arabidopsis* Information Resource *(TAIR)*: gene structure and function annotation. *Nucleic Acids Res.* **36**, D1009–D1014 (2008).
58. Arabidopsis Interactome Mapping Consortium. Evidence for network evolution in an *Arabidopsis* interactome map. *Science* **333**, 601–607 (2011).
59. Mazzucotelli, E. *et al.* The E3 ubiquitin ligase gene family in plants: regulation by degradation. *Curr. Genom.* **7**, 509–522 (2006).
60. Liaw, M.-L. Review of livemocha. *Lang. Learn. Technol.* **15**, 36–40, http://www.lltjournal.org/item/2722 (2011).
61. Zafarani, R. & Liu, H. Social computing data repository at ASU, http://socialcomputing.asu.edu (2009).
62. Dempster, A. P., Laird, N. M. & Rubin, D. B. Maximum likelihood from incomplete data via the EM algorithm. *J. Roy. Stat. Soc. B Met.* 1–38 (1977).
63. Csárdi, G. & Nepusz, T. The igraph software package for complex network research. *InterJournal* Complex Systems, 1695, http://igraph.org (2006).
64. Huang, X.-T., Zhu, Y., Chan, L. L. H., Zhao, Z. & Yan, H. An integrative *C. elegans* protein–protein interaction network with reliability assessment based on a probabilistic graphical model. *Mol. BioSyst.* **12**, 85–92 (2016).
65. Wickham, H. *ggplot2: Elegant Graphics for Data Analysis*, http://ggplot2.org (Springer-Verlag New York, 2009).

## Acknowledgements

The authors are grateful to Dr. Peng Wang for making the source code for the PNet program available. Mark Girolami, Illia Horenko, Ritabrata Dutta, Pavel Krivitsky and Akihiko Nishimura provided helpful discussions on earlier versions of the manuscript. We are grateful to Federica Bianchi for her careful and valuable assistance and advice on earlier drafts. We thank Melbourne Bioinformatics at the University of Melbourne, grant number VR0261, and the Swiss National Supercomputing Centre for high performance computing facilities. M.B. thanks Svitlana Byshkina for her patience. This work was supported by Swiss National Science Foundation grant number 167362, National Research Program 75 ("Big Data") awarded to A. Lomi, and by Swiss National Science Foundation grant number 105218_163196 ("Statistical Inference on Large-Scale Mechanistic Network Models") awarded to A. Mira.

## Author Contributions

M.B., A.S., A.M., G.R. and A.L. conceived the research, designed the research, conducted the analyses and wrote the manuscript. M.B. conceived the original EE algorithm.

## Additional Information

**Supplementary information** accompanies this paper at https://doi.org/10.1038/s41598-018-29725-8.

**Competing Interests:** The authors declare no competing interests.

**Publisher's note:** Springer Nature remains neutral with regard to jurisdictional claims in published maps and institutional affiliations.

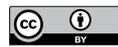 **Open Access** This article is licensed under a Creative Commons Attribution 4.0 International License, which permits use, sharing, adaptation, distribution and reproduction in any medium or format, as long as you give appropriate credit to the original author(s) and the source, provide a link to the Creative Commons license, and indicate if changes were made. The images or other third party material in this article are included in the article's Creative Commons license, unless indicated otherwise in a credit line to the material. If material is not included in the article's Creative Commons license and your intended use is not permitted by statutory regulation or exceeds the permitted use, you will need to obtain permission directly from the copyright holder. To view a copy of this license, visit http://creativecommons.org/licenses/by/4.0/.